\documentclass{aa}
\usepackage[varg]{txfonts}
\usepackage{graphicx}
\usepackage{subcaption}         % for continued figures
\usepackage{lscape}             % to rotate a single page table
\usepackage{placeins}           % provides \FloatBarrier
\usepackage{chapterbib}
\usepackage{titlesec}
\usepackage{float}
\usepackage{threeparttable}
\usepackage{orcidlink}
\usepackage{lineno}
\bibpunct{(}{)}{;}{a}{}{,} % to follow the A&A style
\usepackage{tikz}
\usetikzlibrary{positioning}
\usepackage[percent]{overpic}

\begin{document}

\title{Expansion rate of the young, oxygen-rich supernova remnant G292.0+1.8}

\author{Maria Aslanidou\inst{1}
  \and Manan Agarwal\inst{2,3}\orcidlink{0000-0001-6965-8642}
  \and Jacco Vink\inst{2,3}\orcidlink{0000-0002-4708-4219}
}

\institute{Anton Pannekoek Institute for Astronomy and Astrophysics, University of Amsterdam, Science Park 904, 1098 XH Amsterdam, The Netherlands
  \and Anton Pannekoek Institute/GRAPPA, University of Amsterdam, Science Park 904, 1098 XH Amsterdam, The Netherlands
  \and SRON Netherlands Institute for Space Research, Niels Bohrweg 4, 2333 CA Leiden, The Netherlands}

\date{Received / Accepted}

\abstract{%
Context. Core-collapse supernova remnants (CCSNRs) are ideal targets for studying ejecta–interstellar-medium (ISM) interactions, shock dynamics, and explosion characteristics. The Galactic supernova remnant (SNR) G292.0+1.8 is a classic example of a CCSNR featuring oxygen-rich ejecta, circumstellar material (CSM), a rapidly moving pulsar, and a pulsar wind nebula (PWN).

Aims. This study examines the expansion rate of the Galactic SNR G292.0+1.8 using deep X-ray observations in order to better understand its dynamical evolution and the structure of the ejecta.

Methods. Using deep ACIS-I observations obtained with the Chandra X-ray Observatory over an approximately 10 yr baseline, we measured the X-ray expansion between a 2006 epoch (ObsID 6677) and two 2016 observations (ObsIDs 19892 and 19899), providing two nearly independent baselines (6677–19892 and 6677–19899). After applying relative astrometric corrections based on Gaia DR3 reference sources, we extracted radial profiles in 19 pie-shaped regions around the forward shock and determined the expansion from profile shifts.

Results. The weighted-mean expansion rate is found to be $0.016\% \pm 0.001\%\,\mathrm{yr^{-1}}$ in the broadband. This implies an expansion age of $\sim 2500$ yr for a uniform ambient medium, consistent with previous estimates in the range 2000–3700 yr. For a $1/r^2$ circumstellar density profile, as expected for a Wolf–Rayet progenitor wind, the inferred expansion age is $\sim 4100$ yr. We also analyse three narrow bands dominated by $\alpha$ elements (O–Ne, Mg, and Si–S) and find that lighter elements follow the broadband behaviour, whereas heavier elements show systematically lower expansion rates, consistent with their origin in deeper stellar layers.

Conclusions. Finally, we discuss the pronounced azimuthal asymmetry of the expansion and the apparent paradox that some sectors expand more strongly in the same general direction as the neutron-star kick, as well as the role of reflected shocks driven by the reverse-shock–PWN interaction and the implications for the evolution of the remnant.
}


\maketitle
\nolinenumbers 

\section{Introduction}

Supernova remnants (SNRs) are notable not only for their aesthetic appeal but also for their structural complexity. Young SNRs in our Galaxy and nearby systems, such as the Large Magellanic Cloud, offer unique opportunities to examine the three-dimensional kinematics and chemical composition of the ejecta, an analysis not possible with supernovae (SNe), which appear as unresolved point sources \citep{milisavljevic2017supernova}. The observable features of these young SNRs are shaped by various factors, including the physical processes driving the SN explosion, the mass and composition of the ejecta, and the circumstellar material (CSM) expelled during the progenitor star's late evolutionary phases \citep{patnaude2017impact}.

Supernova remnants therefore offer a valuable tool for examining the structure of the CSM, as well as the asymmetries and elemental composition of SN ejecta. Despite this, establishing clear connections between SNRs and the characteristics of the progenitor stars remains a challenging task. This is particularly true for core-collapse SNRs (CCSNRs) and the mechanisms behind their SN explosions. However, progenitor mass estimates for CCSNRs have been derived, for example through three-dimensional explosion simulations and stellar models that integrate advanced physics. These estimates are further constrained by observational data, such as nitrogen knots and the measured ejecta mass \citep{young2006constraints}, and by abundances of carbon (C), nitrogen (N), and oxygen (O) in shock-heated CSM \citep{uchida2023progenitor}. A well-studied example is Cassiopeia A, an oxygen-rich core-collapse SNR whose ejecta morphology and light-echo spectra reveal a highly asymmetric Type IIb explosion \citep{krause2008cassiopeia,rest11}.
These examples highlight the complexity of SN--SNR connections, which are further influenced by factors such as mass-loss rates, wind speeds, and the evolutionary stage of the progenitor, including possible binary interactions.

One of the most studied Galactic CCSNRs is G292.0+1.8. G292.0+1.8 (hereafter G292) is a young, oxygen-rich SNR first identified in a radio survey by \citet{mills1961catalogue} as MSH~11--5{\it 4}. Its classification as an SNR was later confirmed by \citet{milne19696} based on its non-thermal spectrum. Further confirmation came from \citet{lockhart1977synthesis}, who created a contour map using high-resolution observations from the Fleurs Synthesis Telescope at 1415~MHz. This SNR is believed to be the result of a Type~II core-collapse supernova, as indicated by its distinctive spectrum, which prominently features oxygen and neon lines.

G292 stands out as a compelling target due to its classic characteristics of a CCSNR. It hosts the pulsar PSR~J1124$-$5916 \citep{hughes2001pulsar}, surrounded by a synchrotron-emitting pulsar wind nebula (PWN) \citep{gaensler2003multifrequency}, along with metal-rich ejecta \citep{park2007half} and shocked CSM. X-ray observations further reveal distinct morphologies, including an equatorial belt and thin filaments \citep{park2001structure,park2004nucleosynthesis}.

These unique features make G292 an important object for studying ejecta interactions with the interstellar medium (ISM) and for estimating the progenitor's mass. While multi-wavelength observations across X-rays \citep{park2007half}, the radio \citep{gaensler2003multifrequency}, infrared \citep{lee2009akari}, and optical \citep{winkler2009expanding} have provided valuable insights into its structure and composition, a thorough understanding of its evolutionary history necessitates detailed analysis of its expansion rate.

Calculating the expansion rate of G292 is essential for determining its age, understanding its dynamical evolution, and revealing information about the internal structure of the SNR and the nature of the explosion. The expansion rate provides an upper limit on the age of SNRs such as G292, which undergo deceleration over time, with past estimates in the optical band \citep{ghavamian2005exploring,winkler2009expanding} aligning with the spin-down rate of its central pulsar \citep{camilo2002psr}. Additionally, the observed asymmetries in G292's optical expansion, as shown by the motion of fast-moving ejecta filaments \citep{winkler2009expanding}, suggest anisotropies in the explosion. Similar anisotropies have also been observed in other CCSNRs, such as Cassiopeia~A, through X-ray studies \citep{vink2022forward}.

Despite this, the expansion rate of G292 in X-rays has not yet been investigated. X-rays may either correlate with behaviour seen in other wave bands or probe different hydrodynamical structures. 
Previous X-ray expansion studies of other remnants, such as Tycho’s Type~Ia SNR, highlight the importance of a careful treatment of systematic uncertainties when comparing measurements across wave bands \citep{hughes2000expansion}.

In accurately determining the expansion rate of an SNR such as G292, which is no longer in its earliest stages of evolution,  the ejecta-dominated phase \citep{vink2020physics} presents significant challenges. As the SNR ages, the expansion slows, and the detectable shifts in the ejecta positions over time become more subtle. This makes precise astrometric techniques essential for resolving even small changes in the SNR's structure. Previous X-ray expansion studies of other SNRs, such as Cas A \citep{vink2022forward}, Tycho's SNR \citep{hughes2000expansion,Godinaud2023Tycho3D}, G1.9+0.3 \citep{borkowski2017asymmetric}, and the LMC SNR J0509--6731 \citep{roper2018xray}, have demonstrated the importance of precise astrometric alignment, profile-matching techniques, and the careful treatment of spatially dependent expansion when measuring small shifts.

This paper aims to estimate the average expansion rate of G292 using data from the Chandra X-ray Observatory, specifically from observations conducted in 2006 and 2016. In Sect.~2, we describe the methodology and outline the steps undertaken to derive results for both the broadband and narrow bands. This includes data reprocessing, astrometric corrections, binning, and the application of statistical techniques. Sect.~3 presents the results of the analysis and Sect.~4 discusses their implications for the properties of the SNR, comparing them with findings from previous studies. Finally, Sect.~5 provides a summary of the conclusions and proposes directions for future research.

\section{Observations and data reduction}

\subsection{Data}

The Chandra X-ray data were retrieved using ChaSeR\footnote{\url{https://cda.harvard.edu/chaser/}}, the web interface to the Chandra Data Archive. Two large programmes with an approximately 10~yr baseline were chosen, specifically observations with ObsID~6677 (2006) and ObsIDs~19892 and 19899 (2016), as they offer the longest available time baseline, high exposure times, and similar roll angles, thereby minimising additional systematic uncertainties (Table~\ref{tab:table1}). We selected only ACIS-I datasets to ensure uniform high spatial resolution imaging and downloaded the primary and secondary data products.

\begin{table}
\caption{Chandra observations of SNR G292.0+1.8.}
\label{tab:table1}
\centering
\begin{threeparttable}
\begin{tabular}{l c c c}
\hline\hline
 & ObsID 6677 & ObsID 19892 & ObsID 19899 \\
\hline
Start date & 2006-10-16 & 2016-10-05 & 2016-10-18 \\
Instrument & ACIS-I & ACIS-I & ACIS-I \\
RA (J2000)\tnote{1} & 11:24:39.10 & 11:24:35.65 & 11:24:35.65 \\
Dec (J2000)\tnote{2} & -59:16:20.00 & -59:15:56.38 & -59:15:56.38 \\
Avg.\ count rate\tnote{3} & 34.31 & 21.06 & 21.03 \\
Event count\tnote{4} & 5\,459\,722 & 1\,041\,994 & 895\,189 \\
Exposure (ks) & 159.13 & 49.48 & 42.57 \\
Roll angle (deg) & 140.19 & 150.19 & 144.19 \\
\hline
\end{tabular}
\begin{tablenotes}
    \item[1] Right ascension of the pointing.
    \item[2] Declination of the pointing.
    \item[3] Average count rate in counts per second.
    \item[4] Level~2 event count.
\end{tablenotes}
\end{threeparttable}
\end{table}

Using CIAO version~4.15 and CALDB~4.9.5, we reprocessed each observation with the \texttt{chandra\_repro}\footnote{\url{https://cxc.cfa.harvard.edu/ciao/ahelp/chandra_repro.html}} script to generate new level~2 event lists. This procedure produced reprocessed level~2 event files suitable for subsequent analysis.

\subsection{Astrometric corrections}

Chandra's absolute pointing accuracy is generally better than $0\farcs4$\footnote{\url{https://cxc.cfa.harvard.edu/ciao/threads/reproject_aspect/}}. This can be improved in an absolute sense by cross-matching Chandra sources with higher-precision catalogues, or in a relative sense by cross-matching sources from one Chandra observation with sources from another observation. In this paper, we choose the second approach to improve the relative astrometry between the two observation epochs.

First, we ran the script \texttt{wavdetect}\footnote{\url{https://cxc.cfa.harvard.edu/ciao/threads/wavdetect/}} to detect source candidates. The tool starts by correlating the input dataset with `Mexican-hat' wavelet functions at various scale sizes. We chose scales of 1, 2, and 4 pixels, which are ideal for positional-accuracy studies, especially when the main interest is in the core of the source. \texttt{wavdetect} uses as inputs an exposure-corrected image, a congruent exposure map, and a point-spread function (PSF) map that were made using the script \texttt{fluximage}\footnote{\url{https://cxc.cfa.harvard.edu/ciao/ahelp/fluximage.html}} in the broadband (0.5--5~keV), with the level~2 event files we created as input. Creating exposure-corrected images using an exposure map is beneficial since, in regions with substantial exposure variations, \texttt{wavdetect} may otherwise inaccurately estimate detection significance.
Using an exposure map suppresses false positives and helps refine source-property estimates.

\begin{table}
\caption{Reference source coordinates and Gaia DR3 IDs for observation 19892.}
\label{tab:refsrc_combined}
\centering
\begin{tabular}{c c c c}
\hline\hline
Region & RA (deg) & Dec (deg) & Gaia DR3 ID \\
\hline
1 & 171.10553 & -59.38866 & 5339168201843657088 \\
2 & 171.09743 & -59.30955 & 5339172187529797504 \\
3 & 170.98749 & -59.30242 & 5339172531127097344 \\
4 & 170.94739 & -59.25973 & 5339170372911128320 \\
5 & 170.87929 & -59.23428 & 5339169509123452672 \\
6 & 171.06681 & -59.20067 & 5339174214798279296 \\
\hline
\end{tabular}
\end{table}

Two parameters were required to make the PSF map. The \texttt{eband} (effective energy band) parameter must be the same effective energy used to create the exposure map, namely 2.3~keV, and the \texttt{ecf} (enclosed counts fraction) value is the percentage of the source counts at the given effective energy that is desired to be encompassed by the PSF. This value determines the size of the PSF map that is produced; \texttt{ecf} = 0.393 corresponds to the integrated volume of a 2D Gaussian, encompassing about 39.3\% of the source counts. This is suitable for high-precision localisation and minimising background inclusion.

\begin{figure}[h!]
\centering
 \includegraphics[width=9cm]{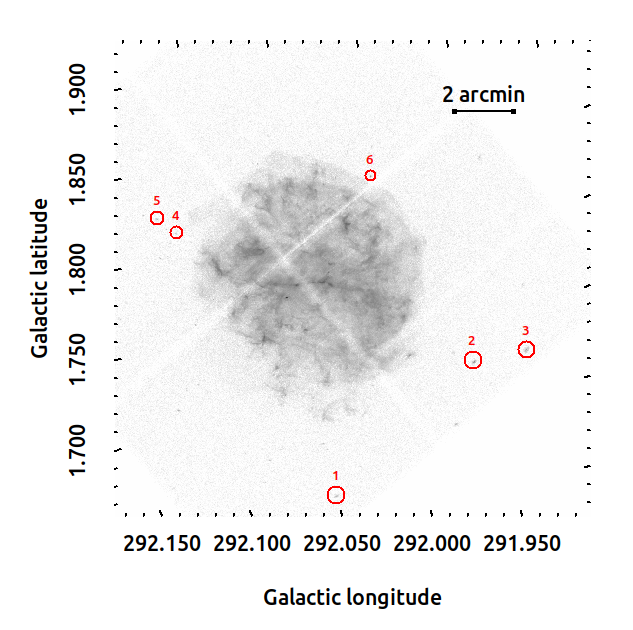}
   \caption{Candidate sources are indicated by red circles for clarity. Shown is the broadband image from ObsID~6677.
   The image orientation follows equatorial (RA–Dec) axes, while the coordinates displayed are in the Galactic coordinate system; the same convention is used throughout the remaining figures.}
  \label{fig:Sources}
\end{figure}

To verify their nature, we cross-matched the detected sources with the Gaia DR3\footnote{\url{https://www.cosmos.esa.int/web/gaia/dr3}} catalogue. DS9 interfaces with this high-precision catalogue, and we found matched sources within a $1\farcs5$ radius. From these matched sources, we selected only those with a high signal-to-noise ratio, to ensure adequate counts for better analysis, and that are not located at the telescope's edge. We identified six candidate sources that cover the whole SNR, as shown in Fig.~\ref{fig:Sources}. Most of these sources were also previously reported by \citet{long2022proper}. Table~\ref{tab:refsrc_combined} lists the Gaia coordinates and identifiers of these six sources.

The CIAO tools \texttt{wcs\_match}\footnote{\url{https://cxc.cfa.harvard.edu/ciao/ahelp/wcs_match.html}} and \texttt{wcs\_update}\footnote{\url{https://cxc.cfa.harvard.edu/ciao/ahelp/wcs_update.html}} were used to compute the fine astrometric shifts between two source lists and apply the offsets to various Chandra files.

\texttt{wcs\_match} calculates a transformation matrix that minimises the error between tangent-plane projections of the reference sources and the transformed input sources. Using the default \texttt{method=rst} (rotate, scale, translate), the transformation matrix specifies a 2D translation, a rotation about the tangent point, and a scale factor. Applying this matrix to the input sources results in the best attempt at aligning the reference and transformed input source positions.

\texttt{wcs\_update} implements the transformation file data by calculating four equivalent parameters to update the \texttt{outfile}: two translational, one rotational, and one scaling parameter. We first updated the aspect solution file and then used this file to further update the event file with respect to our reference file.

The transformation indicates that the second epoch needs to be shifted approximately one pixel to the left and $+0.87$~pixels upwards. In terms of right ascension (RA) and declination (Dec), this corresponds to a slight westward shift and a northward shift. The fitted scale factors were negligible and fully consistent with unity for both ObsID~19892 and ObsID~19899, indicating no measurable relative plate-scale differences between the epochs. This matrix is then applied to the six regions, and the results for both ObsID~19892 and ObsID~19899 are shown together in Table~\ref{tab:wcs_resid_both}, and summarised in Table~\ref{tab:wcs_stats_both}.

\begin{table*}[!h]
\caption{Summary of WCS-matching results -- residuals for ObsIDs~19892 and 19899.}
\label{tab:wcs_resid_both}
\centering
{\small
\begin{tabular}{c c c c c c c}
\hline\hline
Region &
Prior resid$_{92}$ &
Transfm resid$_{92}$ &
Ratio$_{92}$ &
Prior resid$_{99}$ &
Transfm resid$_{99}$ &
Ratio$_{99}$ \\
 & (arcsec) & (arcsec) &  & (arcsec) & (arcsec) &  \\
\hline
1 & 0.88 ( 0.57,$-0.67$) & 0.11 ( 0.11,$-0.03$) & 0.20 &
    0.39 ( 0.35,$-0.18$) & 0.20 ( 0.12, 0.16) & 0.49 \\
2 & 0.47 ( 0.23,$-0.41$) & 0.07 ($-0.03$, 0.06) & 0.32 &
    0.44 ($-0.13$,$-0.43$) & 0.06 ($-0.02$,$-0.06$) & 0.29 \\
3 & 0.45 ( 0.10,$-0.43$) & 0.13 ($-0.13$, 0.01) & 0.51 &
    0.62 ($-0.36$,$-0.51$) & 0.22 ($-0.15$,$-0.16$) & 0.87 \\
4 & 0.87 ( 0.21,$-0.84$) & 0.39 ($-0.05$,$-0.39$) & 0.61 &
    0.96 ( 0.16,$-0.95$) & 0.35 ($-0.15$,$-0.32$) & 0.75 \\
5 & 0.16 ( 0.16,$-0.05$) & 0.41 ($-0.09$, 0.40) & 0.83 &
    0.63 ( 0.33,$-0.53$) & 0.12 ($-0.01$, 0.12) & 0.15 \\
6 & 0.50 ( 0.30,$-0.40$) & 0.20 ( 0.19,$-0.06$) & 0.79 &
    0.35 ( 0.14,$-0.32$) & 0.33 ( 0.20, 0.25) & 1.87 \\
\hline\hline
\end{tabular}
}
\end{table*}

\begin{table*}[ht]
\caption{Summary of WCS-matching results for ObsIDs~19892 and 19899 statistical metrics.}
\label{tab:wcs_stats_both}
\centering
{\small
\begin{tabular}{c c c c c}
\hline\hline
Metric & Before$_{92}$ & After$_{92}$ & Before$_{99}$ & After$_{99}$ \\ \hline
Average residuals (arcsec) &
0.5558 &
0.2183 (60.72\%) &
0.5656 &
0.2136 (62.24\%) \\
Maximum residuals (arcsec) &
0.8815 &
0.4080 (53.71\%) &
0.9639 &
0.3508 (63.60\%) \\
RMS residuals (arcsec) &
0.4313 &
0.1807 (58.10\%) &
0.4259 &
0.1676 (60.66\%) \\
Average residual ratios &
1.5168 &
0.5428 (64.21\%) &
1.7220 &
0.7367 (57.22\%) \\
Maximum residual ratios &
2.0948 &
0.8296 (60.40\%) &
2.4531 &
1.8709 (23.73\%) \\
RMS ratios &
1.1495 &
0.4168 (63.74\%) &
1.2952 &
0.6559 (49.36\%) \\
\hline\hline
\end{tabular}
}
\end{table*}

The results of the Chandra observations after applying the WCS-matching transformations indicate significant improvements in the astrometric accuracy. For ObsID~19892, the average residual decreased from $0\farcs56$ to $0\farcs22$, and for ObsID~19899 from $0\farcs57$ to $0\farcs21$, corresponding to improvements of about 60--63\%. Similar reductions are seen in the maximum and RMS residuals, as well as in the residual ratios (Table~\ref{tab:wcs_stats_both}). These reductions suggest that the applied frame transformations have effectively corrected the positional discrepancies, aligning the observations with the reference frame much more precisely.

\subsection{Radial profiles}

After correcting for astrometric errors, we selected 19 pie-shaped regions within the SNR, avoiding chip gaps that could skew the results. These regions, each spanning about $10^\circ$--$17^\circ$ in azimuth, were chosen to ensure sufficient counts and to accurately capture the shifts. All regions share a common centre, identified as the optical expansion centre by \citet{winkler2009expanding}, with coordinates RA = 11:24:34.4000 and Dec = $-59$:15:51.000 (J2000).

\begin{figure*}[!ht]
\centering
\begin{minipage}{0.32\textwidth}
    \centering
    \includegraphics[width=\linewidth]{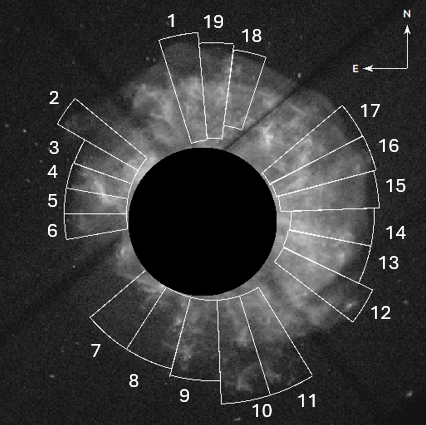}
    \caption*{(a) ObsID 6677}
\end{minipage}
\hfill
\begin{minipage}{0.32\textwidth}
    \centering
    \includegraphics[width=\linewidth]{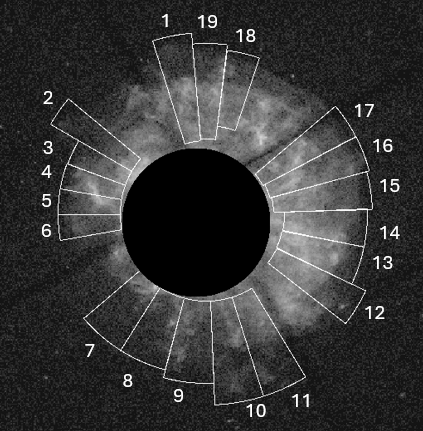}
    \caption*{(b) ObsID 19892}
\end{minipage}
\hfill
\begin{minipage}{0.32\textwidth}
    \centering
    \includegraphics[width=\linewidth]{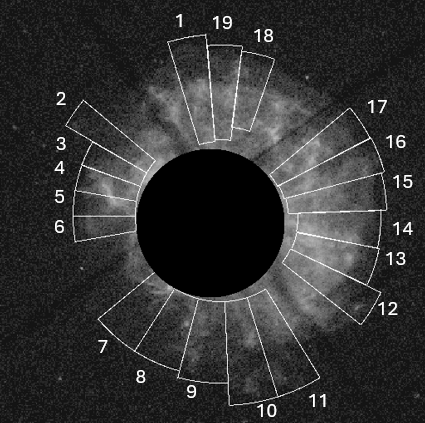}
    \caption*{(c) ObsID 19899}
\end{minipage}

\caption{Selected pie-shaped regions used for the radial profile analysis.
Panels (a)–(c) show the regions overlaid on the Chandra images for
ObsIDs~6677, 19892, and 19899, respectively.}
\label{fig:pie_regions}
\end{figure*}

\begin{figure*}[ht]
\sidecaption
\centering
\begin{overpic}[width=12cm]{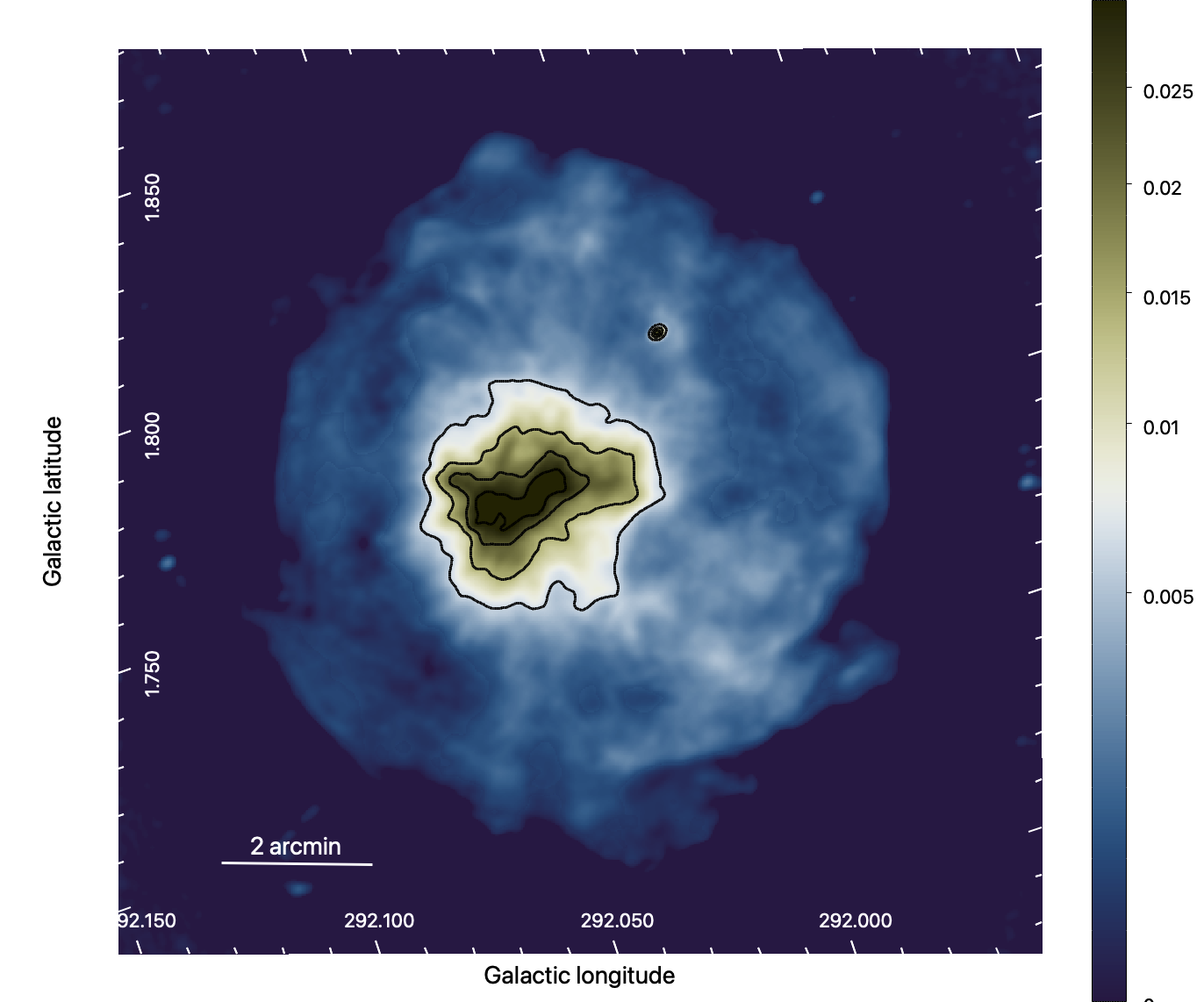}
\put(0,55){
\fbox{\includegraphics[width=3.2cm]{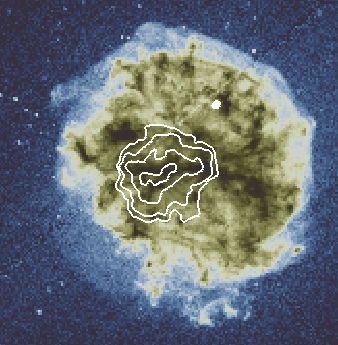}}
}
\end{overpic}
\caption{MeerKAT 1.33 GHz radio map of G292.0+1.8 \citep{cotton24} with added contours to show the extent of the pulsar wind nebula. Contours are square-root spaced from 6 mJy beam$^{-1}$ to 30 mJy beam$^{-1}$. The inset shows an X-ray view of the reverse-shock region for ObsID~6677, highlighting the structure used to define the inner radial boundary. The MeerKAT radio contours are shown on top.}
\label{fig:RadioContours}
\end{figure*}

The reverse-shock region, at a radius of approximately $130''$ \citep{bhalerao2015x}, was excluded to prevent negative shifts and to focus on shifts from the centre to the forward shock, estimated at $265''$ \citep{bhalerao2015x}. Fig.~\ref{fig:pie_regions} illustrates the selected regions, and Fig.~\ref{fig:RadioContours} showcases the radio contours highlighting the interaction with the reverse shock.

We applied a $\chi^2$ calculation with a brightness scaling factor by incrementally shifting the radial midpoint values of the first dataset by $0\farcs1$. For each shift, overlapping bins with the second dataset were identified and included in the $\chi^2$ calculation, adjusted by a scaling factor.

The best shift and scaling factor, which minimise $\chi^2$, were identified by iterating over shifts from $-1''$ to $+1''$ and scaling factors from 0 to 1 in steps of $0\farcs1$ and 0.01, respectively. Data were rebinned by aggregating midpoint and count values over defined intervals, excluding any excess.

To refine the results, a quadratic fit around the minimum $\chi^2$ value was applied to accurately determine the best shift and its 90\% confidence interval, corresponding to $\Delta \chi^2=2.7$. The output includes the optimal shift, the $\chi^2$ at that shift, and the reduced $\chi^2$. Visualisations display $\chi^2$, the fit, and confidence intervals (see Appendix~A).

\subsubsection{Narrow-band analysis}

For the analysis within the narrow energy bands, the C-statistic was employed in conjunction with mild Gaussian smoothing to suppress statistical noise. The Gaussian $\sigma$ ranged from 0.1 to 0.3\,pixels in order not to adversely affect the statistics. We assume that the counts $N_2$ in the second (shallower) epoch are Poisson distributed around the model $\alpha N_1$, where $N_1$ is taken from the deepest observation and treated as the model profile. In this way, only the Poisson error on $N_2$ enters the likelihood. A quadratic fit was similarly used for determining the 90\% confidence intervals.

The C-statistic is defined as \citep{cash79} :
\begin{equation}
C = -2 \ln P
  = -2 \sum_{i,j} \left[ N_{2} \ln \left( \alpha N_{1} \right)
  - \alpha N_{1}
  - \ln \left( N_{2}! \right) \right] ,
\end{equation}
where $N_{2}$ and $N_{1}$ are the counts in profiles~2 and~1, respectively, and $\alpha$ is the normalisation factor. This factor scales the shallower counts so that, in the absence of structural differences, the shape and amplitude of the two profiles can be directly compared.

\section{Results}
\subsection{Broadband}

\begin{figure*}
\centerline{
  \includegraphics[width=0.48\textwidth]{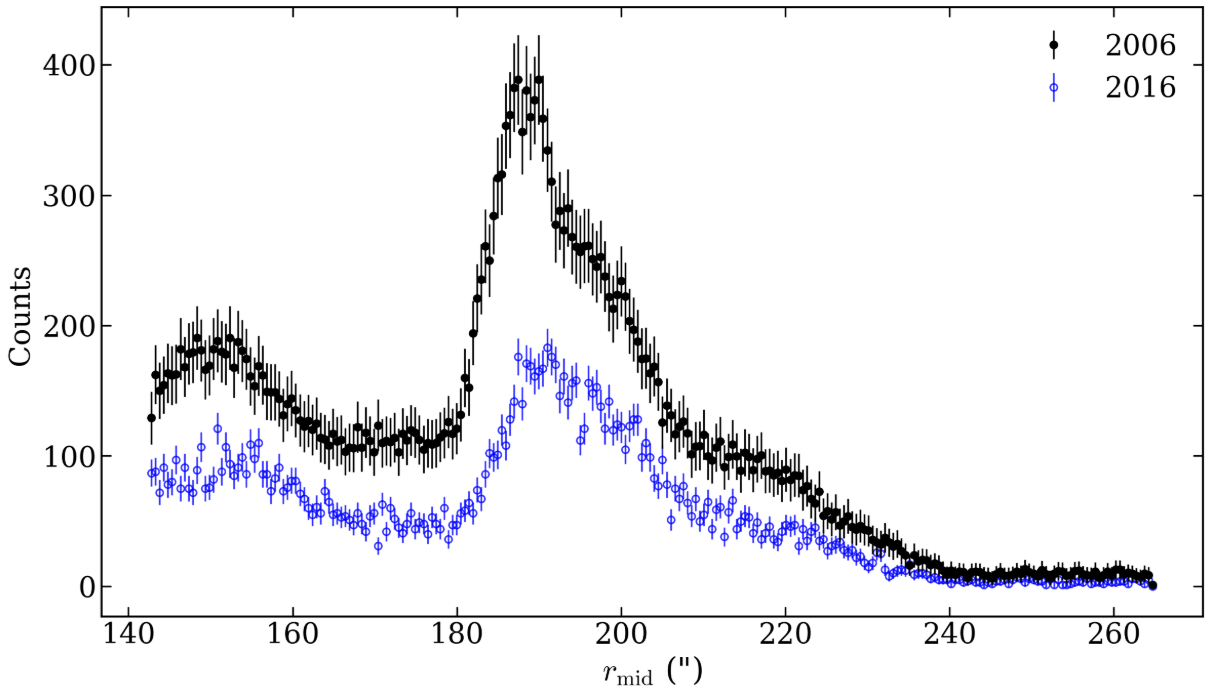}
  \hspace{2mm}
  \includegraphics[width=0.48\textwidth]{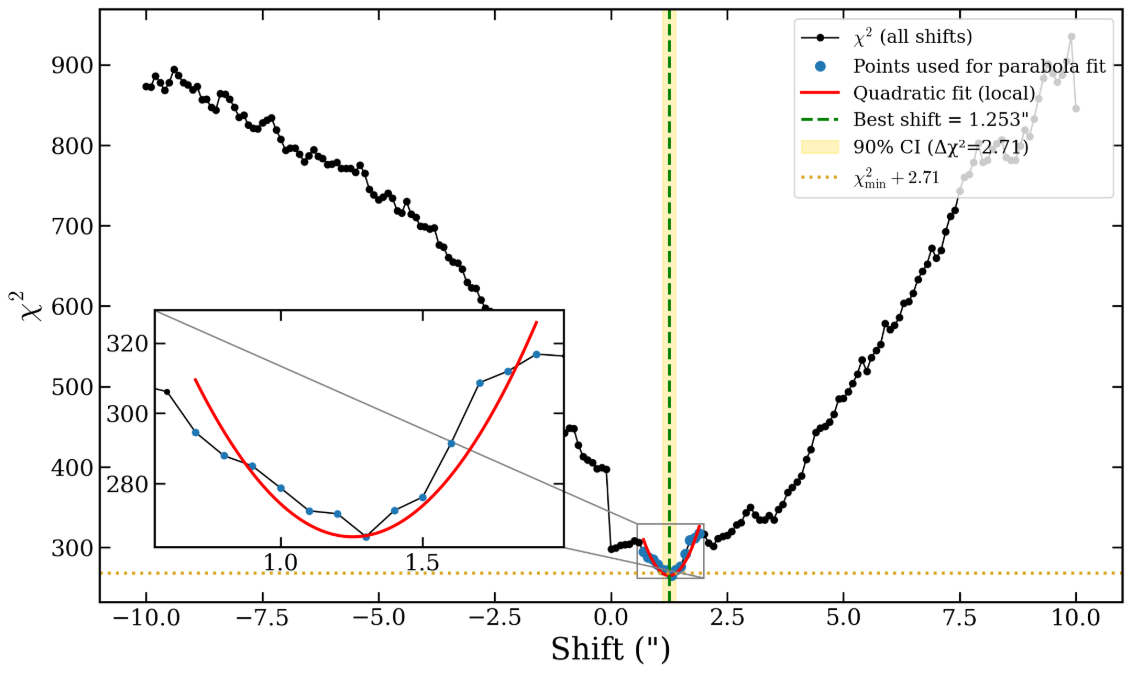}
}
\caption{
Left: Region~5 radial profile.
Right: Region~5 $\chi^2$--shift curve and confidence interval.
}
\label{fig:pixel_e_spectra}
\end{figure*}

Figure~\ref{fig:pixel_e_spectra} presents example radial profiles for the broadband (0.5--5~keV). The error bars reflect the uncertainties in the counts using Gehrels statistics \citep{gehrels86}. Since there is a significant difference in exposure time between the two epochs, the profiles from 2006 have been scaled by a factor of 0.3 for illustration purposes.

By examining the radial profiles, we observe significant shifts in certain regions, such as regions~3, 5, and~6, while in other regions, such as regions~12, 13, and~16, the shifts are barely noticeable. These trends become clearer when we analyse their numerical values.

The shifts derived from the $\chi^2$ minimisation are used to compute proper motions for each region. Since relative astrometric corrections were applied between epochs, we account only for the $0\farcs1$ relative astrometric uncertainty reported by the Chandra calibration team as a systematic contribution to the proper-motion uncertainties.

It is evident that there is azimuthal asymmetry in the inferred motions, with certain areas of the SNR exhibiting larger displacements than others. This asymmetry will be further investigated in Sect.~\ref{sec:asymmetry}.

\subsubsection{Proper motions and expansion rates}

The proper motions were obtained by dividing the measured radial shifts and their associated uncertainties by the temporal baseline between the two observing epochs. These angular motions were converted into transverse velocities using the standard relation
\begin{equation}
v_\perp = 4.74 \times 10^3 \, \mu \, d ,
\end{equation}
where $v_\perp$ is in $\mathrm{km,s^{-1}}$, $\mu$ is the proper motion in $\arcsec,\mathrm{yr^{-1}}$, and $d$ is the distance in kpc. The numerical factor follows from the small-angle relation between angular displacement and physical transverse displacement. A displacement of $1\arcsec$ at a distance of 1 pc corresponds to 1 AU; hence, at a distance of 1 kpc, a proper motion of $1\arcsec\,\mathrm{yr^{-1}}$ corresponds to $1000~\mathrm{AU\,yr^{-1}}$. Since
\begin{equation}
1~\mathrm{AU\,yr^{-1}} =
\frac{1.495978707 \times 10^8~\mathrm{km}}
{3.15576 \times 10^7~\mathrm{s}}
= 4.74047~\mathrm{km\,s^{-1}},
\end{equation}
the conversion becomes
\begin{equation}
v_\perp = 4.74047 \times 10^3 \, \mu \, d
\simeq 4.74 \times 10^3 \, \mu \, d .
\end{equation}
We adopted a distance of $6.2 \pm 0.9$ kpc for G292.0+1.8 \citep{gaensler2003multifrequency}. The resulting velocities therefore correspond only to the plane-of-the-sky component of the shock motion.

The expansion rates of the remnant were then calculated by normalising the proper motions by the radius of the forward shock, providing a direct measure of the fractional radial growth of the remnant per unit time. This approach allows a consistent comparison between different regions and between the two observation epochs.

Table~\ref{tab:pm_exp_combined} summarises, for each region, the position angle, proper motion, transverse velocities, reduced $\chi^2$ of the shift fit, and the corresponding expansion rate for both ObsIDs~19899 and~19892. The reduced $\chi^2$ values indicate that the majority of the fits are statistically acceptable, confirming the reliability of the shift measurements.

Although the measured shifts differ slightly between ObsIDs~19899 and~19892, owing to differences in exposure time and effective temporal baseline, the resulting proper motions and expansion rates are consistent within the quoted uncertainties. This indicates no statistically significant epoch-dependent differences in the inferred expansion velocities and demonstrates the robustness of the astrometric corrections and radial-profile fitting procedure.

Since the expansion rates exhibit clear azimuthal asymmetry across the remnant, a simple arithmetic mean would not adequately represent the global expansion behaviour. We therefore compute a weighted mean expansion rate, where each measurement is weighted by the inverse square of its uncertainty. This approach ensures that measurements with smaller uncertainties contribute more strongly to the final estimate.

The weighted mean expansion rate ($\bar{x}_\mathrm{w}$) is defined as
\begin{equation}
\bar{x}_\mathrm{w} = \frac{\sum \left(x_i / \sigma_i^2\right)}{\sum \left(1 / \sigma_i^2\right)} ,
\label{eq:weighted_mean}
\end{equation}
and its standard error ($\mathrm{SE}_\mathrm{w}$) as
\begin{equation}
\mathrm{SE}_\mathrm{w} = \sqrt{\frac{1}{\sum \left(1 / \sigma_i^2\right)}} .
\label{eq:standard_error}
\end{equation}

Applying this method to the broadband measurements yields a weighted-mean expansion rate of 0.016\%~yr$^{-1}$, with a standard error of 0.001\%~yr$^{-1}$. This value is adopted as the characteristic X-ray expansion rate of G292.0+1.8.
Previous optical studies of G292.0+1.8 measured the proper motions of oxygen-rich 
[O III] filaments and found motions mostly in the range of 
$20$--$100~\mathrm{mas~yr^{-1}}$, with a free-expansion kinematic age of 
$2990 \pm 60$ yr \citep{winkler2009expanding}.
This corresponds to an approximate free-expansion rate of $\sim 0.033\pm 0.001\%~\mathrm{yr^{-1}}$, higher 
than the weighted-mean X-ray expansion rate measured here. However, the two 
measurements trace different physical components: the optical result follows 
fast-moving oxygen-rich ejecta filaments, whereas our X-ray measurement traces 
the broader shocked X-ray-emitting plasma and forward-shock regions.

\subsection{Narrow bands}

\begin{figure}[h!]
    \vspace*{1cm}
    \centering
    \includegraphics[width=0.5\textwidth]{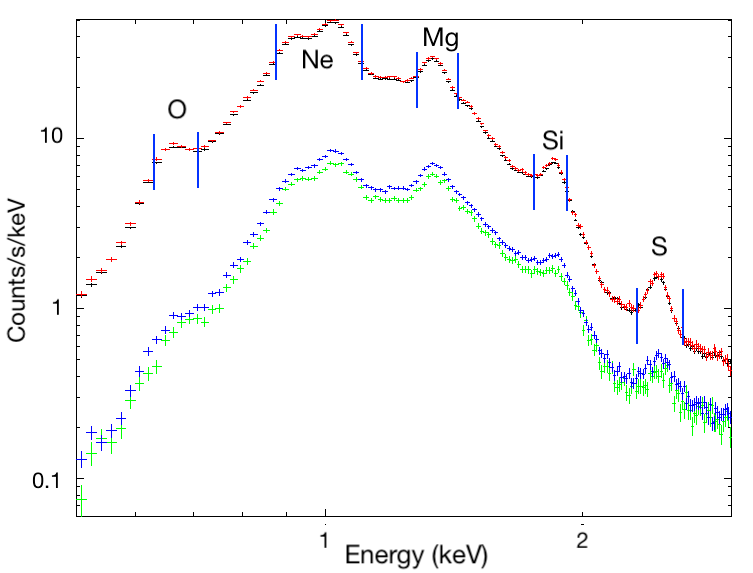}
    \caption{Spectrum of the three observations of SNR~G292.0+1.8 obtained with XSPEC. Vertical lines indicate emission lines from different elements.}
    \label{fig:spectrum}
\end{figure}

Having obtained an initial understanding of the overall behaviour in the broadband, we now proceed to analyse the individual elements. To accurately select the energy bands, we employed HEASoft's XSPEC to extract the integrated X-ray spectrum of the SNR. The result is shown in Fig.~\ref{fig:spectrum}. In the spectrum, we identify five prominent emission lines corresponding to $\alpha$-elements, which are characteristic of a typical core-collapse supernova.

\begin{figure*}[!htb]
\centering

% --- Main tricolour image ---
\includegraphics[width=0.83\textwidth]{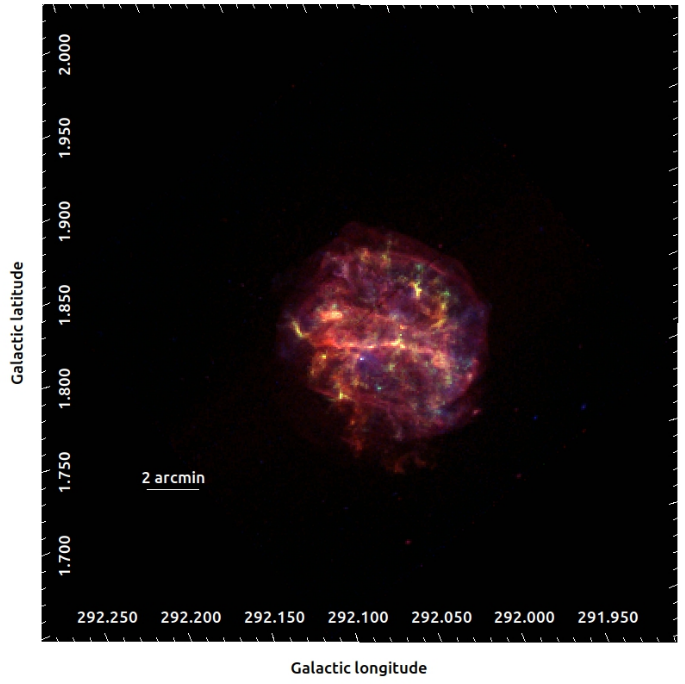}

\vspace{4mm}

% --- Subfigures ---
\begin{subfigure}{0.31\textwidth}
  \centering
  \includegraphics[width=\linewidth]{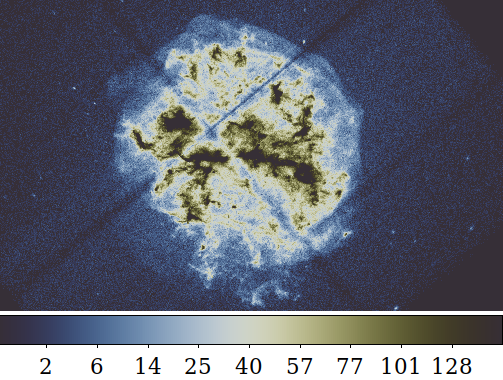}
  \caption{O--Ne}
  \label{fig:gne_on}
\end{subfigure}
\hfill
\begin{subfigure}{0.31\textwidth}
  \centering
  \includegraphics[width=\linewidth]{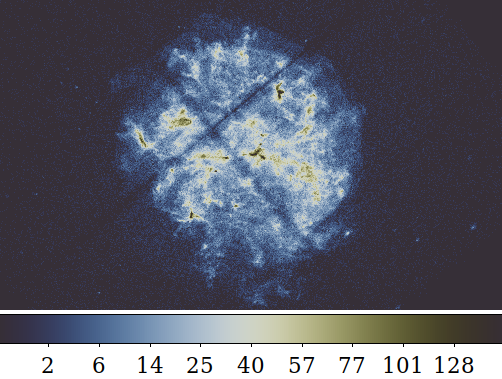}
  \caption{Mg}
  \label{fig:gne_mg}
\end{subfigure}
\hfill
\begin{subfigure}{0.31\textwidth}
  \centering
  \includegraphics[width=\linewidth]{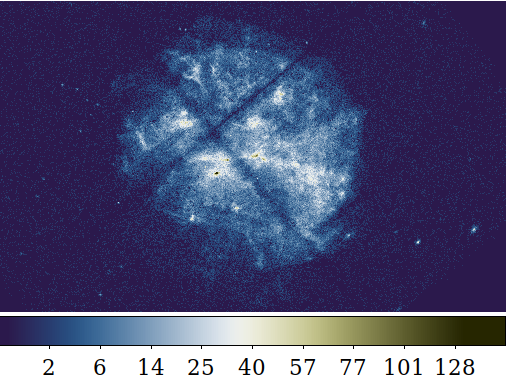}
  \caption{Si--S}
  \label{fig:gne_sis}
\end{subfigure}

\caption{
Tricolour Chandra image of SNR~G292.0+1.8 from ObsID~6677.
Red corresponds to O--Ne, green to Mg, and blue to Si--S.
The panels below show the individual narrow-band images used to construct the tricolour map:
(a) O--Ne band, 0.58--0.71~keV (O Ly$\alpha$) and 0.88--0.95~keV (Ne He$\alpha$);
(b) Mg band, 1.28--1.43~keV (Mg He$\alpha$);
(c) Si--S band, 1.81--2.05~keV (Si He$\alpha$) and 2.40--2.62~keV (S He$\alpha$).
All images are smoothed for display purposes.
}
\label{fig:figure3.5}
\end{figure*}

To ensure sufficient counts for statistical analysis, we combined neighbouring energy ranges into three broader narrow-band intervals: an O--Ne band (0.58--0.71~keV for O~Ly$\alpha$ and 0.88--0.95~keV for Ne~He$\alpha$), an Mg band (1.28--1.43~keV, Mg~He$\alpha$), and a Si--S band (1.81--2.05~keV for Si~He$\alpha$ and 2.40--2.62~keV for S~He$\alpha$). To visualise the spatial distribution of these elements, we created a tricolour flux image in DS9, with a small Gaussian smoothing applied to enhance the visibility of faint structures (Fig.~\ref{fig:figure3.5}).

Figure~\ref{fig:figure3.5} illustrates the spatial distribution and count density of these elements around the SNR. In the O--Ne band, the emission is widely dispersed with ample counts, indicating a broad presence throughout the remnant. Magnesium exhibits a more limited distribution with fewer counts, suggesting a less widespread presence. The Si--S emission is more centrally concentrated, with higher counts in the inner regions and fewer counts at larger radii.

We calculated the shifts, proper motions, and expansion rates for these three narrow energy bands using the same procedure as for the broadband. Owing to the lower counts in the narrow bands, we employed the C-statistic instead of the $\chi^2$ statistic and applied mild Gaussian smoothing with $\sigma = 0.1$--0.3\,pixels to mitigate noise and occasional zero-count bins without significantly biasing the profiles.

\subsubsection{Oxygen--neon}

Oxygen (O) and neon (Ne) are among the most abundant elements produced during the late stages of stellar evolution in massive stars that undergo core-collapse supernovae. They are synthesised in the outer layers of the progenitor star, particularly in the helium- and carbon-burning shells. As a consequence of their origin in these relatively external layers, O and Ne tend to be widely dispersed throughout the remnant after the explosion.

Consistent with this picture, the O--Ne narrow-band profiles closely follow the broadband behaviour. The inferred shifts and proper motions are very similar to those obtained in the broadband, and the resulting expansion rate closely matches the broad-band value. The weighted-mean expansion rate for the O--Ne band is $0.016\%$~yr$^{-1}$, with a standard error of $0.001\%$~yr$^{-1}$.

\subsubsection{Magnesium}

Magnesium (Mg) is synthesised in layers closer to the stellar core during the later stages of evolution, typically in the oxygen-burning shell, which lies deeper than the regions where O and Ne are formed. As a result, the Mg-emitting material is expected to exhibit somewhat lower expansion velocities and a more confined spatial distribution than the O--Ne band.

This expectation is borne out by our measurements: the Mg-band shifts and proper motions are slightly smaller than those seen in the broadband, and the weighted-mean expansion rate is correspondingly reduced to $0.014\%$~yr$^{-1}$, with a standard error of $0.001\%$~yr$^{-1}$.

\subsubsection{Silicon--sulfur}

Silicon (Si) and sulfur (S) are synthesised in the innermost layers of the progenitor during the final stages of nuclear burning, within the silicon-burning shell just outside the iron core. Because they originate deep in the star, the Si- and S-rich ejecta are expected to undergo less extensive outward mixing and to have lower expansion velocities than the outer layers traced by O and Ne.

This is reflected in the observed Si--S narrow-band profiles: the emission is more centrally concentrated, and the measured shifts and proper motions are systematically lower than those in the broadband and in the lighter elements. The weighted-mean expansion rate for the Si--S band is $0.011\%$~yr$^{-1}$, with a standard error of $0.001\%$~yr$^{-1}$.

Before turning to the discussion, we summarise the broad-band and narrow-band expansion rates in Table~\ref{tab:expansion_rates}.

\begin{table}
\caption{Expansion rates for the broadband and the three narrow bands.}
\label{tab:expansion_rates}
\centering
\begin{tabular}{l c c}
\hline\hline
Energy band & Expansion rate (\% yr$^{-1}$) & Error (\% yr$^{-1}$) \\
\hline
Broadband & 0.0160 & 0.0010 \\
O--Ne      & 0.0161 & 0.0012 \\
Mg         & 0.0142 & 0.0010 \\
Si--S      & 0.0113 & 0.0005 \\
\hline\hline
\end{tabular}
\end{table}

\section{Discussion}
\label{sec:discussion}

\subsection{Age}

The age of an SNR can be estimated using the relation
\begin{equation}
\mathrm{age} = \beta \times (\mathrm{expansion\ rate})^{-1} ,
\end{equation}
where $\beta$ is the expansion parameter. The Sedov--Taylor solution \citep{sedov2018similarity} provides a theoretical framework for understanding the expansion dynamics of an SNR in a medium with a power-law density profile $\rho(r) \propto r^{-s}$. This is a good approximation once the swept-up mass exceeds the ejecta mass and the inner ejecta are reached by the reverse shock. In this context, the radius $R_\mathrm{s}$ of the shock wave and its velocity $V_\mathrm{s}$ can be described as
\begin{equation}
R_\mathrm{s} \propto t^\beta, \quad V_\mathrm{s} = \beta \frac{R_\mathrm{s}}{t} ,
\end{equation}
with the expansion parameter given by
\begin{equation}
\beta = \frac{2}{5 - s} .
\end{equation}

\begin{itemize}
\item For $s = 0$, the Sedov--Taylor solution assumes that the explosion energy $E$ is instantaneously injected into a uniform medium with constant density $\rho_0$ (i.e. a point explosion), and that there are no energy losses. This scenario corresponds to the classic Sedov phase where the expansion parameter is
    \begin{equation}
    \beta = \frac{2}{5} .
    \end{equation}

\item For $s = 2$, the scenario changes to an SN shock moving through the progenitor's stellar wind, which has a density profile decreasing with the square of the radius. This is a more astrophysically relevant case, particularly for young SNRs like Cassiopeia~A \citep{vink2012supernova}. G292.0 might also be in this regime, as supported by studies such as \citet{temim2022snr}. Here, the expansion parameter becomes
    \begin{equation}
    \beta = \frac{2}{3} .
    \end{equation}
\end{itemize}

Using the estimated broadband expansion rate of $0.016\%$~yr$^{-1}$ and accounting for an error of $0.001\%$~yr$^{-1}$, we can calculate the expansion age of the SNR for both scenarios.

\begin{itemize}
    \item For $s = 0$ (uniform medium),
    \begin{equation}
        \mathrm{Age} \approx 2500~\mathrm{yr} ,
    \end{equation}
    with a plausible range of $\sim2300$--2800~yr when accounting for the uncertainty on the expansion rate.

    \item For $s = 2$ (stellar-wind density profile),
    \begin{equation}
        \mathrm{Age} \approx 4160~\mathrm{yr} ,
    \end{equation}
    with a corresponding range of $\sim3800$--4700~yr.
\end{itemize}

The spin-down age of the pulsar, $\sim 2900$ yr \citep{camilo2002psr}, and the optical kinematic age of $2990 \pm 60$ yr derived from [O III] filament proper motions \citep{winkler2009expanding}, lie between the two idealised expansion-age estimates obtained above. They are closer to the uniform-medium estimate than to the simple $s=2$ wind-profile estimate. If the optical or pulsar age is combined with our measured X-ray expansion rate of $0.016\%~\mathrm{yr}^{-1}$, the implied expansion parameter is $\beta \simeq 0.48$, intermediate between the Sedov value for a uniform medium, $\beta=0.4$, and the value for a steady wind profile, $\beta=2/3$. This suggests that the dynamical evolution of G292.0+1.8 is unlikely to be described by either idealised density profile alone, and that the circumstellar environment may have a more complex structure.

Moreover, \citet{long2022proper}, who measured the proper motion of the pulsar, indicated that the characteristic age might be closer to $2000$~yr. This lower estimate still falls within the combined error bounds of the expansion-age calculations, particularly when considering the lower-luminosity, wind-like scenario.

It is important to note that there may be discrepancies between the pulsar's true age $t$ and its characteristic age $\tau$. While the characteristic age is often used as an estimate of the pulsar's age, it can differ significantly from the true age depending on the pulsar's spin-down history and braking index, which vary with different radiation models. This is notable considering that for many pulsars, their characteristic ages are much older than their associated SNRs. For example, PSR~B1706$-$44 has a characteristic age of about 17\,500~yr, while its associated SNR G343.1$-$2.3 is only about 5000~yr old \citep{nicastro1996evidence,abramowski2011detection}.

Finally, the expansion-age results for G292 are also consistent with previous estimates. Specifically, the age range of approximately 2700--3700~yr inferred by \citet{chevalier2005young} aligns well with our typical case calculation. Furthermore, the age derived from optical filaments was estimated to be $2990 \pm 60$~yr \citep{winkler2009expanding}, which also lies between the two expansion ages inferred here. Additionally, this optical filament age aligns closely with the spin-down age of the central pulsar, which was also estimated to be around 2900~yr. The consistency between the expansion age and the spin-down age reinforces the reliability of these measurements and our understanding of the SNR's evolutionary stage. Taken together, the clustering of independent age estimates around $\sim2900$~yr suggests that the X-ray expansion results may favour an expansion parameter $\beta < 0.5$ rather than the $\beta = 2/3$ expected for a steady wind density profile ($s=2$). However, if G292 originated from a massive progenitor with a complex mass-loss history, the circumstellar density structure may deviate from that of a simple steady wind, in which case Eq.~(9) may not strictly apply.

\subsection{Asymmetry of the explosion}
\label{sec:asymmetry}

Observational evidence supporting an asymmetric SN explosion for G292 has been reported in previous works: higher proper motions of optical ejecta knots along the north--south than in the east--west directions \citep{winkler2009expanding} and a $\sim1$~pc displacement (to the southeast of the SNR's expansion centre) of the associated pulsar PSR~J1124$-$5916 \citep{winkler2009expanding}.

\begin{figure*}[!ht]
    \vspace{0.5cm}
    \centering
    \includegraphics[width=0.9\textwidth]{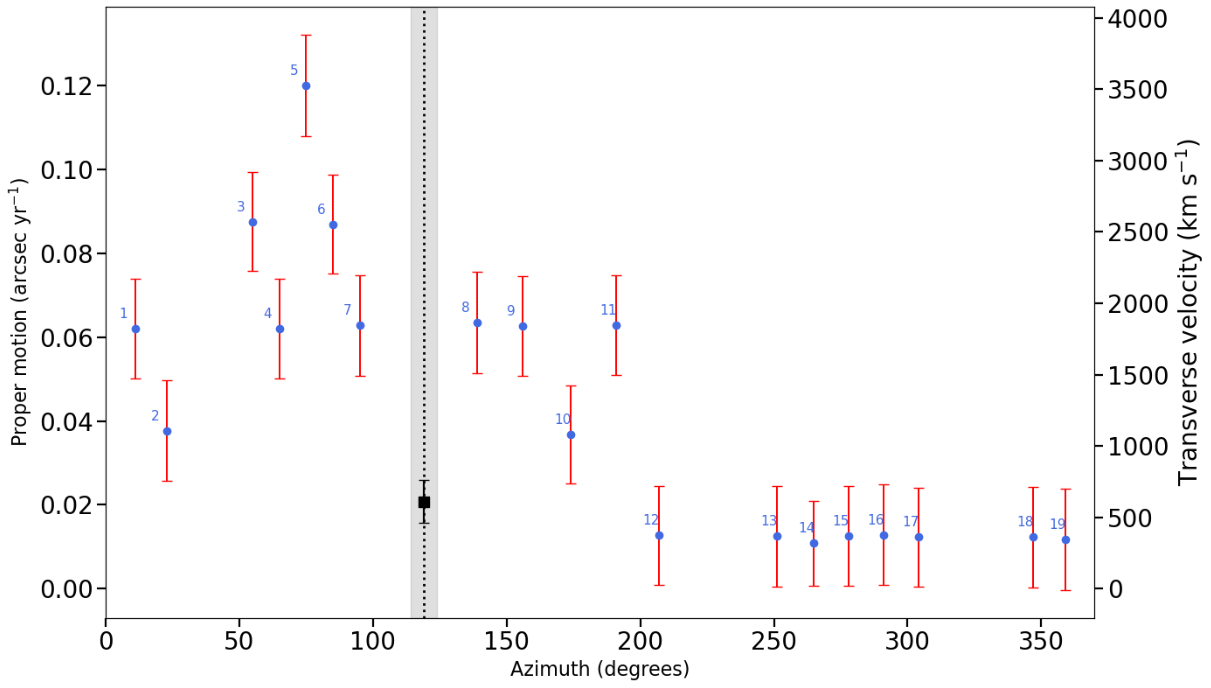}
    \caption{Proper motion versus position angle for the broadband.
Blue circles show the transverse velocities of the X-ray--emitting ejecta knots with their associated uncertainties.
The black square marks the transverse velocity of the neutron star, $v_\perp \approx 612 \pm 152$~km~s$^{-1}$, at a position angle of $119^\circ \pm 5^\circ$ \citep{long2022proper}.
The dotted vertical line indicates the inferred kick direction of the neutron star, while the shaded region represents the uncertainty on its position angle.
}
    \label{fig:figure3.1}
\end{figure*}

As previously noted, there is a significant difference in the shifts observed between the western and eastern sides of the SNR. This asymmetry is apparent in both the broadband and all the narrow bands analysed. To further investigate this, we plotted position angle versus proper motion to examine their behaviour around the remnant, as shown in Figs.~\ref{fig:figure3.1} and \ref{fig:figure3.2}, along with the pulsar's velocity and direction.

\begin{figure*}[!htbp]
\centering
\begin{minipage}{0.47\textwidth}
    \centering
    \includegraphics[width=\linewidth]{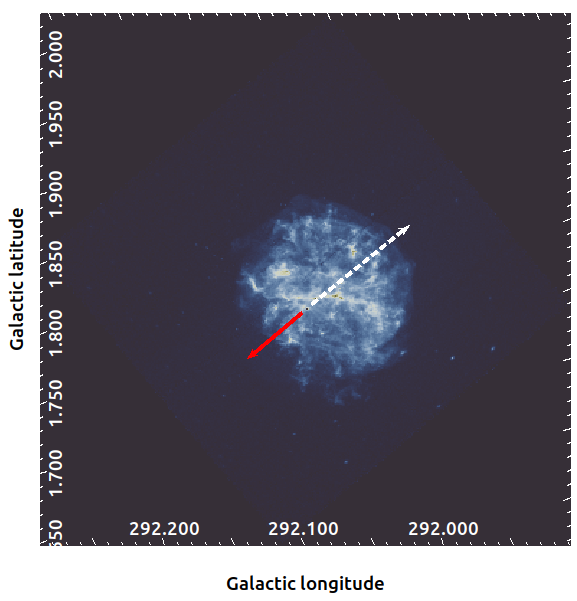}
    \subcaption{Dipole moment and neutron-star motion.}
    \label{fig:dipole_ns}
\end{minipage}
\hfill
\begin{minipage}{0.48\textwidth}
    \centering
    \includegraphics[width=\linewidth]{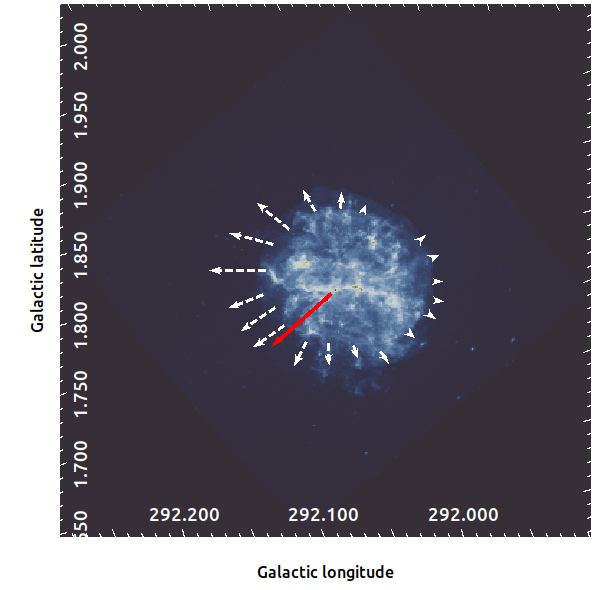}
    \subcaption{Illustration of azimuthal proper motions and neutron-star motion.}
    \label{fig:pm_arrows}
\end{minipage}
\caption{
Chandra 0.5--7~keV images of ObsID~6677 illustrating asymmetries in the explosion and expansion.
Left: The white arrow points from the explosion centre towards the direction of the dipole moment, while the red arrow indicates the direction of the neutron-star motion.
Right: Schematic illustration of the measured proper motions around the remnant, showing that the expansion is systematically larger in the direction opposite to the neutron-star motion. Arrow lengths are indicative and not to scale.
}
\label{fig:asymmetry_overview}
\end{figure*}

An important observation is that only when the reverse shock begins compressing the PWN from the direction of the pulsar’s motion does the pulsar become displaced from the geometric centre of the PWN towards the east (Fig.~\ref{fig:dipole_ns}). The left panel illustrates the large-scale asymmetry of the system: the direction of the neutron-star motion (red arrow) is misaligned with the ejecta dipole moment (white arrow), a configuration expected from an intrinsically asymmetric explosion and consistent with the neutron-star kick mechanism discussed by \citet{holland2017comparing}. In such scenarios, conservation of momentum would suggest that the bulk of the ejecta expands more rapidly in the direction opposite to the neutron-star motion.

As shown in the right panel (Fig.~\ref{fig:pm_arrows}), the expansion is systematically larger in the same general direction as the neutron-star motion, rather than opposite to it. This indicates that the observed asymmetry in the expansion cannot be explained solely by the initial explosion geometry. Instead, it suggests that later-time dynamical effects—such as anisotropic interaction of the reverse shock with the PWN and ejecta, reflected shocks, and a non-uniform ambient density—play a significant role in shaping the present-day expansion pattern. Together, the eastward displacement of the pulsar within the PWN and the direction-dependent expansion rates measured in Fig.~\ref{fig:pm_arrows} can therefore account for part of the unexpected asymmetry observed in the SNR’s measured shifts. 

At first glance, this might seem counter-intuitive. One plausible scenario to explain this phenomenon is that the ejecta are being impacted by a reflected shock. According to \citet{temim2022snr}, at an age of $\sim2500$~yr, the majority of the PWN's surface has been reached by the reverse shock. \citet{gaensler2003multifrequency} also support this statement by conducting a multifrequency radio study of G292. In their work, they argue that the PWN is in the early stages of interaction with the SNR reverse shock; it is possible that this process produces a complicated magnetic-field geometry.\footnote{For a detailed analysis, see Sect.~4.3 of \citet{gaensler2003multifrequency}.} This indicates that most of the supernova ejecta have been shocked at this stage. As the ejecta and reverse shock collide, a reflected shock forms and propagates away from the explosion centre, moving through the large-scale filaments of ejecta in the southeast.

\subsection{Shock dynamics}
\label{sec:shock_dynamics}

The expansion rates observed in the O--Ne bands are very close to those seen in the broadband, indicating similar dynamic behaviour within the SNR. This suggests that these elements, which reside in the outer layers of the ejecta, are expanding at comparable rates due to similar interactions with the surrounding environment.

In contrast, the Mg band, and most prominently the Si--S band, exhibit lower expansion rates. This discrepancy may arise because Mg, Si, and S are located in the inner layers of the SNR. These elements are more significantly decelerated by the surrounding material and by the reverse shock. The increased deceleration is likely a consequence of the higher density in the inner layers, which impedes the expansion of these elements more than those in the outer layers.

The interaction of the SN shock wave with denser regions of the ISM or CSM significantly affects the dynamics of the SNR. The progenitor star’s wind or mass-loss episodes play a crucial role in creating a denser circumstellar environment. These episodes result in regions of varying density around the SNR, which subsequently interact differently with the shock wave and ejecta. A denser CSM can decelerate the inner layers of the SNR more effectively, while the outer layers experience less deceleration, leading to the observed differences in expansion rates and shifts between the elements.

When the shock wave from an SNR encounters a denser medium, it experiences increased resistance, causing the ejecta (the material expelled by the SN) to slow down more significantly. This greater deceleration in denser regions also leads to more homogeneous mixing of elements within the SNR. As a result, the elements, despite their initial velocity differences, start interacting with the surrounding medium in a more similar manner. Regions that initially travelled faster will decelerate significantly when encountering these dense areas, resulting in slower observed shifts.

In the case of the northwestern part of the SNR (approximately between 250$^\circ$ and 350$^\circ$ in PA), the slower shifts can be attributed to this interaction with a denser medium. These regions may have encountered denser material earlier, causing them to decelerate more compared to other parts, such as the northeast. Initially, the northwestern regions may have travelled at higher speeds, reaching and interacting with the dense medium sooner, leading to more significant deceleration over time.

Another factor contributing to the observed lower shifts is the projection of three-dimensional velocities onto the plane of the sky. The true velocities of the ejecta in the SNR are three-dimensional, but we can only observe the component of these velocities projected onto the plane of the sky (the two-dimensional view from Earth). Variations in observed shifts can result from the angle at which these velocities are projected. Different parts of the SNR may appear to move more slowly simply because their velocity vectors have a larger component directed towards or away from us (along the line of sight) rather than across the plane of the sky.

\section{Conclusions}

This study provides the first estimation of the X-ray expansion rate for the SNR~G292, revealing a rate of $0.016\%\pm0.001\%$~yr$^{-1}$, derived from a decade of observations (2006 and 2016) using two nearly independent baselines (6677--19892 and 6677--19899). The calculated expansion age lies between approximately 2500 and 4200~yr, which is broadly consistent with previous estimates based on optical filament proper motions and the central pulsar's spin-down age \citep{winkler2009expanding,camilo2002psr}.

Significant azimuthal variations in expansion were observed, with the eastern portion expanding more than other regions. This discrepancy is attributed to the interaction between the PWN and the reverse shock, creating a reflected shock that further impacts the surrounding ejecta, consistent with previous studies \citep{temim2022snr}. Notably, this introduces an apparent paradox: in some sectors, the largest expansion is observed broadly in the same direction as the neutron-star kick, rather than opposite to it as a simple momentum argument might suggest.

Lower shifts in certain parts of the remnant’s forward shock are likely due to interactions with denser ISM or CSM, especially opposite the pulsar's kick direction. Initially high-velocity ejecta encountered dense material, slowing down and indicating a complex dynamical interplay between the remnant’s material and its environment.

Lighter elements, such as oxygen and neon, follow the broadband expansion pattern, reflecting the SNR's overall dynamics. Heavier elements, such as magnesium, silicon, and sulfur, show lower velocity shifts due to stronger deceleration by the reverse shock and the CSM. Nevertheless, they behave similarly to the lighter elements in regions with higher broadband shifts, indicating that the ejecta structure and shock interactions are coherent on large scales.

Future work should include additional observational epochs, more detailed simulations of PWN--reverse-shock--ISM interactions, and extended multi-wavelength campaigns to better understand the SNR's dynamic evolution and the link between explosion geometry and the surrounding medium.

\begin{acknowledgements}
The research on this project by MA and JV was (partially) funded by NWO under grant number 184.034.002.
\end{acknowledgements}

\bibliographystyle{aa}
\bibliography{mybib}

\FloatBarrier
\clearpage

\begin{appendix}

\section{Radial-profile extraction and implementation details}

Our Python code performs radial-profile analysis of photon events stored in FITS files. The analysis is based on event lists and is designed to measure small radial shifts between epochs with subpixel accuracy.

Using the \texttt{astropy} library, the world coordinate system (WCS) is initialised for each dataset, allowing transformation between pixel and celestial coordinates. Photon events are then extracted within user-defined annular sectors centred on the expansion centre of the remnant.

\subsection{Event selection and geometry}

Annular sectors are defined by the following parameters:
\begin{itemize}
    \item $x_\mathrm{centre}, y_\mathrm{centre}$: centre coordinates,
    \item $r_1, r_2$: inner and outer radii of the annulus,
    \item $\theta_\mathrm{min}, \theta_\mathrm{max}$: angular bounds in degrees.
\end{itemize}

For each event $i$, the coordinate differences relative to the centre are computed as
\begin{equation}
x_\mathrm{diff} = x_i - x_\mathrm{center}, \quad
y_\mathrm{diff} = y_i - y_\mathrm{center} .
\end{equation}

The corresponding radius and position angle are given by
\begin{equation}
r_i = \sqrt{x_\mathrm{diff}^2 + y_\mathrm{diff}^2} ,
\end{equation}
\begin{equation}
\theta_i =
\left(
\arctan2(y_\mathrm{diff}, x_\mathrm{diff})
\frac{180}{\pi}
\right)
\bmod 360 .
\end{equation}

Events are selected if they satisfy
\begin{equation}
(r_i \geq r_1) \land (r_i \leq r_2) \land
(\theta_i \in [\theta_\mathrm{min}, \theta_\mathrm{max}]) .
\end{equation}

\subsection{Binning and radial profiles}

In Chandra data analysis, binning is used to improve the signal-to-noise ratio. The native ACIS pixel size of $0\farcs5$ was oversampled by a factor of five, yielding an effective radial sampling of $0\farcs1$. This approach enables accurate localisation of the forward shock while preserving sensitivity to small shifts.

Radial profiles are generated using the extracted right ascension and declination values. The angular distance of each event from the centre is computed as
\begin{align}
\mathrm{ra}_\mathrm{diff} & =
(\mathrm{ra} - \mathrm{ra}_\mathrm{center})
\cos(\mathrm{radians}(\mathrm{dec}_\mathrm{center})), 
\\
\mathrm{dec}_\mathrm{diff} & =
(\mathrm{dec} - \mathrm{dec}_\mathrm{center}) ,
\\
\mathrm{radii}_\mathrm{distance} & =
\sqrt{\mathrm{ra}_\mathrm{diff}^2 + \mathrm{dec}_\mathrm{diff}^2} .
\end{align}

The number of radial bins is defined as
\begin{equation}
N_\mathrm{bins} =
\left\lfloor (r_2 - r_1) \times \mathrm{factor} \right\rfloor ,
\end{equation}
where \texttt{factor} is the oversampling factor. Bin edges are computed incrementally, and the midpoint of each bin (\texttt{rmid}) is defined as the mean of adjacent edges.

\subsection{Uncertainties}

Count-rate uncertainties are computed using Gehrels statistics \citep{gehrels86}:
\begin{equation}
\mathrm{counts}_\mathrm{err} =
1 + \sqrt{\mathrm{counts} + 0.75} .
\end{equation}

This prescription is appropriate for the low-count regime encountered in the narrow-band analysis.

Figure~\ref{fig:pie_regions} in the main text illustrates the spatial subdivision of the remnant into annular sectors following this procedure.

\section{Broad-band proper motions and expansion rates}
\label{app:broad_band_table}

This appendix provides the full set of broadband proper-motion and expansion-rate measurements for the individual regions used in the radial-profile analysis. The table~\ref{tab:pm_exp_combined} lists the results separately for the two 2016 observations, ObsIDs~19899 and~19892, using ObsID~6677 as the reference epoch. These measurements form the basis for the weighted-mean broad-band expansion rate discussed in Sect.~3.1.

\begin{table*}
\caption{Broad-band proper motions, expansion rates, and transverse velocities by region for ObsIDs~19899 and~19892.}
\label{tab:pm_exp_combined}
\centering
{\small
\begin{tabular}{c c c c c c}
\hline\hline
ROI & PA ($^{\circ}$)$^{a}$ &
Proper motion ($''$ yr$^{-1}$) &
Expansion rate (\% yr$^{-1}$)$^{b}$ &
$v_{\perp}$ (km s$^{-1}$)$^{c}$ &
$\chi^2/\mathrm{dof}$ \\
\hline
\multicolumn{6}{c}{ObsID 19899} \\
\hline
1  & 11  & $0.062 \pm 0.012$ & $0.023 \pm 0.005$ & $1820 \pm 440$ & 1.051 \\
2  & 23  & $0.038 \pm 0.012$ & $0.014 \pm 0.005$ & $1120 \pm 390$ & 0.612 \\
3  & 55  & $0.087 \pm 0.012$ & $0.033 \pm 0.005$ & $2560 \pm 510$ & 0.897 \\
4  & 65  & $0.062 \pm 0.012$ & $0.023 \pm 0.005$ & $1820 \pm 440$ & 0.564 \\
5  & 75  & $0.125 \pm 0.018$ & $0.047 \pm 0.007$ & $3670 \pm 750$ & 0.965 \\
6  & 85  & $0.087 \pm 0.012$ & $0.033 \pm 0.005$ & $2560 \pm 510$ & 0.920 \\
7  & 95  & $0.063 \pm 0.012$ & $0.024 \pm 0.005$ & $1850 \pm 440$ & 0.801 \\
8  & 139 & $0.064 \pm 0.012$ & $0.024 \pm 0.005$ & $1880 \pm 450$ & 1.005 \\
9  & 156 & $0.063 \pm 0.012$ & $0.024 \pm 0.005$ & $1850 \pm 440$ & 0.665 \\
10 & 174 & $0.037 \pm 0.012$ & $0.014 \pm 0.005$ & $1090 \pm 390$ & 0.703 \\
11 & 191 & $0.063 \pm 0.012$ & $0.024 \pm 0.005$ & $1850 \pm 440$ & 0.947 \\
12 & 207 & $0.013 \pm 0.012$ & $0.005 \pm 0.005$ & $380 \pm 360$ & 1.338 \\
13 & 251 & $0.013 \pm 0.012$ & $0.005 \pm 0.005$ & $380 \pm 360$ & 0.775 \\
14 & 265 & $0.011 \pm 0.010$ & $0.004 \pm 0.004$ & $320 \pm 300$ & 0.888 \\
15 & 278 & $0.013 \pm 0.012$ & $0.005 \pm 0.005$ & $380 \pm 360$ & 0.634 \\
16 & 291 & $0.013 \pm 0.012$ & $0.005 \pm 0.005$ & $380 \pm 360$ & 0.667 \\
17 & 304 & $0.012 \pm 0.012$ & $0.005 \pm 0.005$ & $350 \pm 360$ & 0.888 \\
18 & 347 & $0.012 \pm 0.012$ & $0.005 \pm 0.005$ & $350 \pm 360$ & 0.968 \\
19 & 359 & $0.012 \pm 0.012$ & $0.005 \pm 0.005$ & $350 \pm 360$ & 0.754 \\
\hline
\multicolumn{6}{c}{ObsID 19892} \\
\hline
1  & 11  & $0.061 \pm 0.011$ & $0.023 \pm 0.004$ & $1790 \pm 420$ & 0.982 \\
2  & 23  & $0.037 \pm 0.011$ & $0.014 \pm 0.004$ & $1090 \pm 360$ & 0.589 \\
3  & 55  & $0.086 \pm 0.011$ & $0.032 \pm 0.004$ & $2530 \pm 490$ & 0.842 \\
4  & 65  & $0.061 \pm 0.011$ & $0.023 \pm 0.004$ & $1790 \pm 420$ & 0.531 \\
5  & 75  & $0.126 \pm 0.019$ & $0.048 \pm 0.007$ & $3700 \pm 780$ & 0.911 \\
6  & 85  & $0.085 \pm 0.010$ & $0.032 \pm 0.004$ & $2500 \pm 470$ & 0.873 \\
7  & 95  & $0.062 \pm 0.011$ & $0.023 \pm 0.004$ & $1820 \pm 420$ & 0.761 \\
8  & 139 & $0.062 \pm 0.011$ & $0.023 \pm 0.004$ & $1820 \pm 420$ & 0.953 \\
9  & 156 & $0.061 \pm 0.011$ & $0.023 \pm 0.004$ & $1790 \pm 420$ & 0.631 \\
10 & 174 & $0.036 \pm 0.011$ & $0.014 \pm 0.004$ & $1060 \pm 360$ & 0.674 \\
11 & 191 & $0.062 \pm 0.011$ & $0.023 \pm 0.004$ & $1820 \pm 420$ & 0.899 \\
12 & 207 & $0.012 \pm 0.011$ & $0.005 \pm 0.004$ & $350 \pm 330$ & 1.281 \\
13 & 251 & $0.012 \pm 0.011$ & $0.005 \pm 0.004$ & $350 \pm 330$ & 0.740 \\
14 & 265 & $0.010 \pm 0.009$ & $0.004 \pm 0.003$ & $290 \pm 270$ & 0.842 \\
15 & 278 & $0.012 \pm 0.011$ & $0.005 \pm 0.004$ & $350 \pm 330$ & 0.603 \\
16 & 291 & $0.012 \pm 0.011$ & $0.005 \pm 0.004$ & $350 \pm 330$ & 0.638 \\
17 & 304 & $0.012 \pm 0.011$ & $0.005 \pm 0.004$ & $350 \pm 330$ & 0.851 \\
18 & 347 & $0.012 \pm 0.011$ & $0.005 \pm 0.004$ & $350 \pm 330$ & 0.929 \\
19 & 359 & $0.011 \pm 0.011$ & $0.004 \pm 0.004$ & $320 \pm 330$ & 0.723 \\
\hline\hline
\end{tabular}
}
\tablefoot{
$^{a}$Position angle (PA) in degrees, measured east of north. \\
$^{b}$Expansion rates are computed by normalising the proper motions by a forward-shock radius of $265''$. \\
$^{c}$Transverse velocities are computed using $v_\perp = 4.74047 \times 10^3 \mu d$, with $d = 6.2 \pm 0.9$~kpc.
}
\end{table*}

\section{Supplementary figures}

This appendix presents additional figures used to illustrate the azimuthal dependence of the measured proper motions in the narrow energy bands.

\begin{figure*}[h!]
    \centering
    \begin{minipage}{0.67\textwidth}
        \centering
        \includegraphics[width=\linewidth]{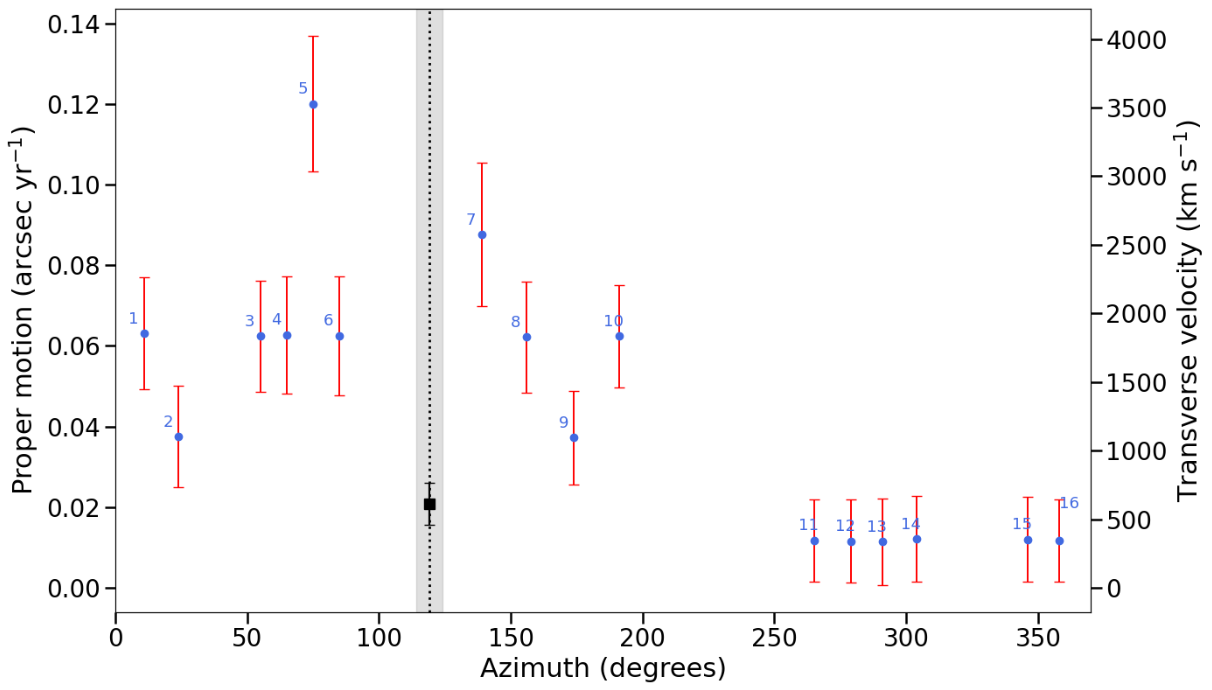}
        \subcaption{Oxygen--neon}
        \label{fig:subfigONe}
    \end{minipage}

    \vspace{5mm}

    \begin{minipage}{0.67\textwidth}
        \centering
        \includegraphics[width=\linewidth]{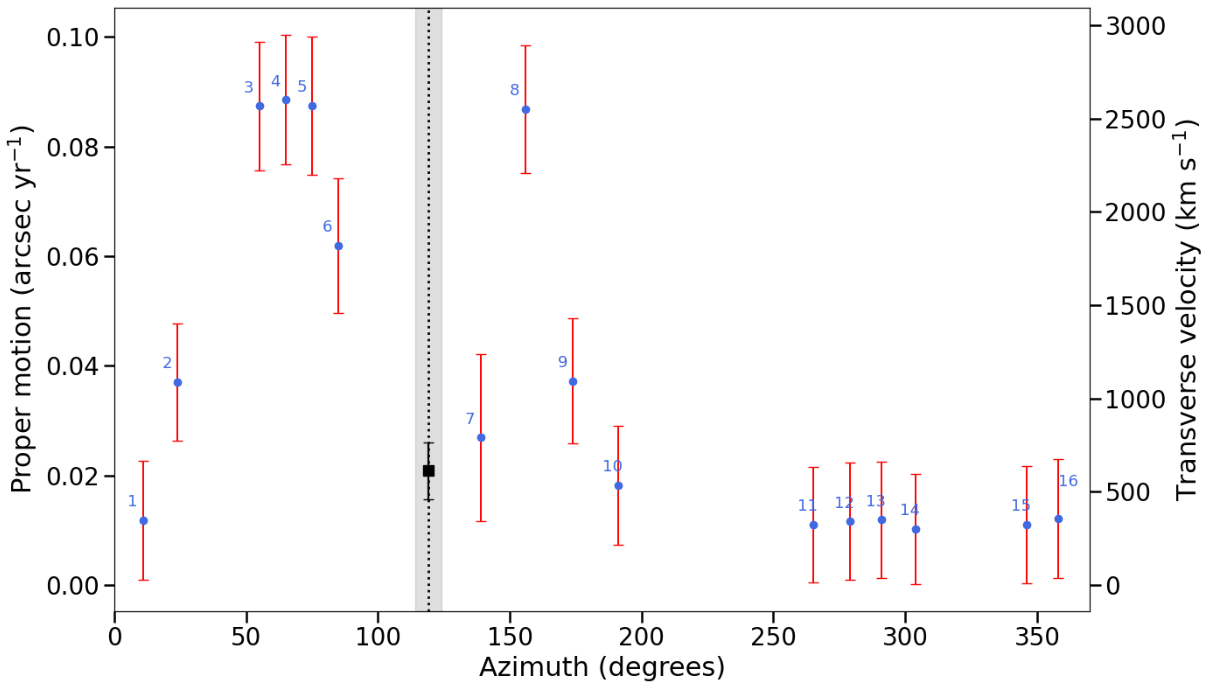}
        \subcaption{Magnesium}
        \label{fig:subfigMg}
    \end{minipage}

    \vspace{5mm}

    \begin{minipage}{0.67\textwidth}
        \centering
        \includegraphics[width=\linewidth]{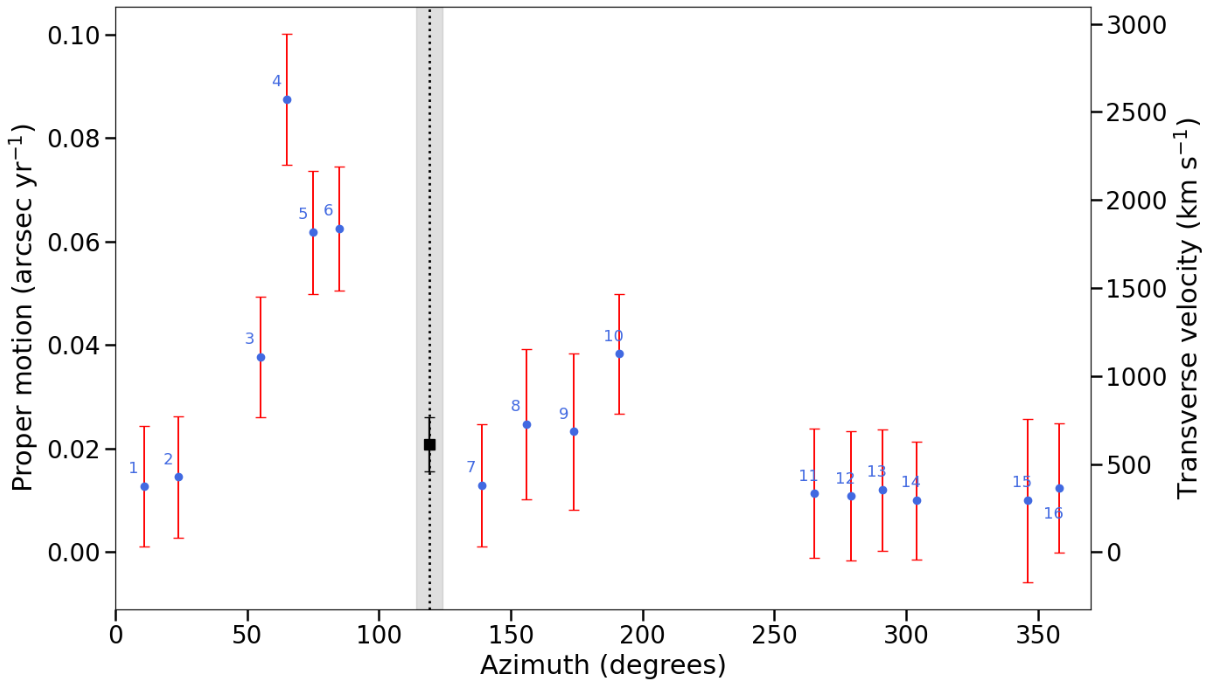}
        \subcaption{Silicon--sulfur}
        \label{fig:subfigSiS}
    \end{minipage}

    \caption{
    Proper motion as a function of position angle for the three narrow energy bands.
    Each panel shows the azimuthal dependence of the measured proper motions.
    }
    \label{fig:figure3.2}
\end{figure*}

Figure~\ref{fig:figure3.2} shows proper motion as a function of position angle for the O--Ne, Mg, and Si--S bands. These plots complement the broad-band results presented in Sect.~3 and demonstrate that the azimuthal trends persist across different elements, while heavier elements systematically exhibit lower expansion velocities. Although the broad-band analysis was performed using 19 regions, not all of these regions were retained in the narrow-band measurements. Some regions were excluded because the narrow-band images contain fewer counts than the broad-band data, leading to poor statistics and poorly constrained radial-profile shifts. In these cases, the C-statistic curves did not provide a reliable minimum, and the derived proper motions would therefore be dominated by statistical noise rather than by a robust physical displacement.

\end{appendix}

\end{document}